\documentclass[aps,prd,11pt,nofootinbib,nobibnotes,eqsecnum,secnumarabic,superscriptaddress,tightenlines,notitlepage,preprintnumbers]{revtex4-1}
\pdfoutput=1

\usepackage{amsmath, amssymb, graphicx}
\usepackage{palatino}
\usepackage{xspace}
\usepackage{braket}
\usepackage{color}
\usepackage{enumitem}
\usepackage[lofdepth,lotdepth,caption=false]{subfig}

\setlist{noitemsep, topsep=.25em, labelindent=1.5\parindent, leftmargin=*}

\DeclareMathOperator{\tr}{\mathrm{tr}}

\DeclareMathOperator{\spn}{\mathrm{span}}

\newcommand\dg{\dagger}

\newcommand{\beq}{\begin{equation}}
\newcommand{\eeq}{\end{equation}}

\newcommand{\veps}{\varepsilon}
\renewcommand\Re{\operatorname{Re}}



\begin{document}
\preprint{IMSc/2011/09/11}
\title{Qubit Models of Black Hole Evaporation}
\author{Steven G. Avery}
\email{savery@imsc.res.in}
\affiliation{The Institute of Mathematical Sciences, CIT Campus, Taramani, Chennai, India 600113}
\begin{abstract}
  Recently, several simple quantum mechanical toy models of black hole
  evaporation have appeared in the literature attempting to illuminate
  the black hole information paradox. We present a general class of
  models that is large enough to describe both unitary and nonunitary
  evaporation, and study a few specific examples to clarify some
  potential confusions regarding recent results.  We also generalize
  Mathur's bound on small corrections to black hole dynamics.
  Conclusions are then drawn about the requirements for unitary
  evaporation of black holes in this class of models. We present a
  one-parameter family of models that continuously deforms nonunitary
  Hawking evaporation into a unitary process. The required deformation
  is large.
\end{abstract}
\maketitle

\section{Introduction}

Hawking's calculation of black hole evaporation~\cite{Hawking:1974sw}
leads to a direct conflict between general relativity and the unitary
evolution of quantum mechanics~\cite{Hawking:1976ra}. This prompted
the suggestion that the requirement of unitary evolution should be
relaxed~\cite{Hawking:1982dj, Page:1979tc}, and pure states allowed to
evolve into mixed states. Such nonunitary evolution, however, seems
problematic~\cite{Banks:1983by}.\footnote{See~\cite{Unruh:1995gn,
    Unruh:2012vd}, however, for criticisms of the arguments made
  in~\cite{Banks:1983by}.}

Moreover, since black hole evaporation is a very slow process
involving a large number of emitted particles, and since one only
expects to start recovering information after about one half of the
radiation has been emitted~\cite{Page:1993df}, one might imagine that
unitarity is restored by the accumulation of small (say,
nonperturbative) corrections over the course of the entire evolution
process, cf.~\cite{Page:1993wv}. To evaluate this claim,
Mathur~\cite{Mathur:2009hf} introduces a qubit model of black hole
evaporation and then derives bounds on the entanglement entropy of the
emitted radiation, which show that this scenario is \emph{not
  possible}. Below we generalize his result to consider more general
deformations of the pair creation dynamics. In particular, for small
changes in the evaporation process the entanglement entropy continues
to increase, therefore (barring remnants and variants thereof) the
evolution is not unitary.

Thus, for unitarity to be restored, one needs to make \emph{large}
corrections to the semi-classical evaporation process described by
Hawking. Recently, several specific unitary models have been
introduced~\cite{Czech:2011wy, Giddings:2011ks, Mathur:2011wg} as
proposed alternatives to Hawking's semiclassical evolution. As
discussed in~\cite{Mathur:2011uj}, models of this kind could be called
``burning paper'' models that involve large corrections to the Hawking
evolution. While these models are unitary, they are typically written
in a way that makes it difficult to compare to the semiclassical
evolution and to see how Mathur's bound operates. The difficulty
arises because the bound is derived in terms of dynamics in an
ever-enlarging Hilbert space, whereas unitary models are typically
written as dynamics in a fixed-dimensional Hilbert space.

The primary goal of this paper is to clarify the meaning of Mathur's
bound in~\cite{Mathur:2009hf}. In particular, until one writes unitary
models as (large) corrections to nonunitary, semiclassical evolution,
the meaning of Mathur's result remains obscure. Previous
investigations of corrections in this context either directly
introduce unitary models that are difficult to compare to Hawking
evaporation~\cite{Czech:2011wy, Giddings:2011ks, Mathur:2011wg}; or
consider small corrections to Hawking evaporation that \emph{do not
  produce unitary evolution when made sufficiently
  large~\cite{Mathur:2011wg, Mathur:2010kx}}, and thus are
unconvincing illustrations of the result in~\cite{Mathur:2009hf}. A
second goal is to characterize what kinds of corrections produce the
desired unitary evolution. That the corrections need to be large is a
necessary condition derived in~\cite{Mathur:2009hf}, but sufficient
conditions have not been discussed in the same way.

In order to address the above points, it is necessary to introduce a
general model space that provides a uniform language to discuss both
unitary and nonunitary black hole evaporation.  This allows us, for
example, to continuously deform the semiclassical Hawking evolution to
unitary evolution. One can then explicitly see that the deformation is
large in an appropriate sense, and therefore in agreement with a suitable
generalization of Mathur's argument.  Let us emphasize that it is not
our intention here to advocate for nonunitary evolution, only to
demonstrate that unitarity demands there be a significant
alteration of the traditional semiclassical evolution. To restrict
ourselves to unitary evolution at this stage would be to beg the
question.

In Section~\ref{sec:gen}, we introduce a very general framework for
qubit models of black hole evaporation that is appropriate for both
unitary and nonunitary evolution. Most of the discussion focuses on
how to interpret the models, and their connection to ideas in quantum
information theory. In Section~\ref{sec:models}, we apply the
formalism to a number of sample models; some of the models were chosen
because they were discussed previously in the literature, and others
because they illustrate some interesting issues. In
Section~\ref{sec:bound}, we briefly review Mathur's argument against
small corrections restoring unitary evolution, and generalize the main
entanglement entropy bound to allow arbitary deformations. Mathur's
original result~\cite{Mathur:2009hf} only explicitly considers one
kind of perturbation. In Section~\ref{sec:unitarity}, using
observations from Section~\ref{sec:models}, we discuss what conditions
ensure unitary evolution. In Section~\ref{sec:one-par}, we present a
one-parameter family of models that continuously connects Hawking
evaporation to a unitary model; one sees clearly that the deformation
required is large. In Section~\ref{sec:conc}, we conclude with some
brief comments.

\section{The General Model}\label{sec:gen}

Before presenting our model, we make a few preliminary comments on
nonunitary evolution in Section~\ref{sec:nonunitary}. Then in
Section~\ref{sec:quops}, using some results from quantum information
theory, we explain how to describe nonunitary evolution that still has
a good probabilistic interpretation. Along the way, we clarify some
potential confusions regarding previous work. Finally, we present our
general class of models in Section~\ref{sec:themodel}, discussing the
physical interpretation in Section~\ref{sec:physics}.

\subsection{Nonunitarity}\label{sec:nonunitary}

When we model the evolution of a closed quantum system, the state of
the system is given by a ket $\ket{\psi(t)}$ that satisfies
\begin{equation}\label{eq:unit-evol-1}
\ket{\psi(t)} = U(t)\ket{\psi(0)}
\end{equation}
for a unitary time evolution operator $U(t)$. We may equivalently
write the state of the system as a density matrix $\rho(t) =
\ket{\psi(t)}\bra{\psi(t)}$ where $\rho(t)$ satisfies
\begin{equation}\label{eq:unit-evol-rho}
\rho(t) = U\rho(0)U^\dg.
\end{equation}
Unitary evolution satisfies several nice conditions,
namely,\footnote{We do not claim that these are completely
  independent.}
\begin{enumerate}
\item\label{it:lin} Linearity
\item\label{it:norm} Preservation of the norm: unit norm states evolve
  to unit norm states, ensuring that a probabilistic interpretation
  makes sense. For density matrices, the desired condition is that
  unit-trace, completely positive density matrices evolve to
  unit-trace, completely positive density matrices.
\item\label{it:inv} Invertibility: previous states can be found from the current state.
\item\label{it:pure} Purity: pure states evolve to pure states.
\end{enumerate}
However, it has sometimes been suggested~\cite{Hawking:1982dj,
  Hawking:1976ra, Page:1979tc} that theories of quantum gravity
(especially in the presence of black holes) will not be unitary. Let
us note that the negation of unitary is ambiguous, since it is not
clear which of the above conditions is relaxed. Let us consider three
illustrative mappings.
\begin{enumerate}
\item
This evolution does not conserve probability---the norm is not
conserved; however, pure states still evolve to pure states, and the
evolution is invertible:
\begin{equation}
\ket{\psi} \mapsto \frac{3}{4}\ket{\psi}.
\end{equation}
This kind of evolution is sometimes useful in modelling a system that
decays into something that is outside the model. For instance, in
modeling alpha decay. An equivalent way to describe this evolution is
to say that the system has energies with an imaginary part.  For a
fundamental description that includes all degrees of freedom, however,
the nonconservation of probability is nonsensical.
\item This evolution preserves the norm, is invertible, but evolves
  pure states to mixed states:
\begin{equation}
\ket{\psi_1} \mapsto \rho_1 = \frac{1}{2}\ket{\psi_1}\bra{\psi_1} 
                          + \frac{1}{2}\ket{\phi_1}\bra{\phi_1}\qquad
\ket{\psi_2} \mapsto \rho_2 = \frac{1}{2}\ket{\psi_2}\bra{\psi_2} 
                          + \frac{1}{2}\ket{\phi_2}\bra{\phi_2}\qquad
\dots
\end{equation}
While this model evolves pure states to mixed states, information is
still preserved (in a weak sense),\footnote{Provided large ensembles
  of identical copies of the system, one can with some confidence
  distinguish distinct density matrices; however, this usage of the
  phrase ``information preservation'' is not canonical, and is
  problematic if one starts considering density matrices which are
  very close to each other.  This would mean in an experiment repeated
  many times with identical initial conditions, one could reconstruct
  some information about the initial state from the final state. In a
  technical sense, however, quantum information is lost.  We make the
  distinction here, since the semiclassical description implies that
  information is not preserved even in this weak sense.} and
probability is conserved.
\item
This evolution conserves probability, evolves pure states to pure
states, but is not invertible.
\begin{equation}
\ket{\psi_1} \mapsto \ket{\psi_0}\qquad
\ket{\psi_2} \mapsto \ket{\psi_0}\qquad
\dots
\end{equation}
In this model one cannot reconstruct the past from the current state,
and therefore information is not preserved. In
Section~\ref{sec:G1-broken}, we introduce some models of this type.
\end{enumerate}

In the black hole information paradox, one considers some initial
configuration of matter $\ket{\psi_m}$ that collapses into a black
hole, and then completely evaporates to radiation in a thermal mixed
state $\rho \sim e^{-\beta H}$. Since the final state is both mixed
and independent of the initial state, the evolution \emph{both} fails
to be invertibile and pure in the above senses.

Suppose for the nonce that Hawking's original argument is
correct~\cite{Hawking:1976ra}, and the fundamental theory of quantum
gravity is not unitary. We can no longer write evolution as
in~\eqref{eq:unit-evol-rho}. We restrict our considerations to
evolution that satisfies conditions~\ref{it:lin} and~\ref{it:norm},
but not necessarily condition~\ref{it:inv} and~\ref{it:pure}. Then,
assuming some very basic conditions that ensure a good probabilistic
interpretation, we can write the most general possible evolution in
the operator-sum representation~(cf.~\cite{nielsen})
\begin{equation}\label{eq:op-sum-rep}
\rho(t) = \sum_k E_k \rho(0) E_k^\dg
\end{equation}
for some set of operator $E_k$ that satisfy the completeness relation
\begin{equation}
\sum_k E_k(t)^\dg E_k(t) = I.
\end{equation}
This is one way to write the evolution of an open quantum system, and
the transformation from $\rho(0)$ to $\rho(f)$ is called a ``quantum
operation'' in the quantum information theory
literature~\cite{nielsen}. The operators $E_k$ determine the evolution
of the density matrix. When there is only one $E_k$, the evolution is
unitary.\footnote{Note that we model evolution with discrete time
  evolution, and do not discuss continuous evolution, which might be
  governed by the Lindblad equation.}

\subsection{Quantum Operations}\label{sec:quops}

As in~\cite{Mathur:2009hf, Giddings:2011ks, Mathur:2011uj,
  Mathur:2010kx, Mathur:2011wg}, we model the Hawking evaporation
process as a discrete set of mappings on qubits. In the initial state,
the system consists entirely of matter in a pure state. There have
been some suggestions~\cite{Braunstein:2009my} in this context that
the entanglement between the initial black hole-forming matter and the
outside matter plays an important role; we do not address these issues
at this time.  The initial state is modeled as a set of $n$ ``matter
qubits'':
\begin{equation}
\rho_0 = \ket{\psi_0}\bra{\psi_0} \qquad \ket{\psi_0} \in\spn \left\{\ket{\hat{q}_1\hat{q}_2\cdots\hat{q}_{n}}\right\},
\end{equation}
where each $\hat{q}$ is a qubit, a quantum state labeled by $0$ or
$1$.  After a sequence of intermediate steps, the end state consists
entirely of radiation (again, we are assuming no remnants), modeled as
a (possibly mixed) density matrix acting on $n$ ``radiation'' qubits,
$\rho_f$. Throughout the evolution, we keep the total dimension of the
Hilbert space fixed. This is certainly true for the unitary evolution
of closed systems, but here we put it in as a reasonable assumption.
We are motivated in part by the black hole's entropy. The black hole
initially has $S\sim M^2$, which entirely radiates away on the time
scale $\sim M^3$ with an emission every $\sim M$. This is consistent
with a model having a fixed number of physical qubits.

Following~\cite{Giddings:2011ks}, we use hats to distinguish the
internal black hole qubits from the external radiation qubits. The
hatted qubits represent all degrees of freedom that are inaccessible
outside the black hole; unlike~\cite{Mathur:2009hf}, we do not
distinguish between degrees of freedom from the initial matter, from
gravitational interactions, or any that arise during the evaporation
process. We write basis elements for the final state as
\begin{equation}
  \{\ket{q_nq_{n-1}\cdots q_1}\},
\end{equation}
where we have put the labels on the qubits in reverse order for
reasons which should become clear. At the $i$th step, we have a
density matrix acting on $n-i$ black hole qubits and $i$ radiation
qubits, so that the total dimension of the Hilbert space is fixed. The
evaporation concludes on the $n$th step when there is only radiation:
\begin{equation}
\rho_0 \to \rho_1 \to \cdots \to \rho_{n-1} \to \rho_n
\end{equation}

In general the mapping from $\rho_0$ to $\rho_n = \rho_f$ should be a
quantum operation, which is the composition of $n$ quantum operations
(one for each emission).  Therefore the total evolution from $\rho_0$
to $\rho_n$ (and each intermediate step) may be written in terms of
some $E_k$s, like in Equation~\eqref{eq:op-sum-rep}. Quantum
operations, however, may be written in an equivalent, alternative form
that is more natural when discussing black hole evaporation. This form
connects more directly with the discussion in~\cite{Mathur:2009hf}.

We motivate this alternate form, by noting that any mixed density
matrix may be ``purified'' by enlarging the Hilbert space. For
example, a density matrix of the form
\begin{equation}
\rho = p_1 \ket{A}\bra{A} + p_2 \ket{B}\bra{B}
\end{equation}
with orthonormal $\{\ket{A}, \ket{B}\}$ can be purified by
introducing the orthonormal states $\{\ket{\alpha},\ket{\beta}\}$, and
defining
\begin{equation}
  \ket{\Psi} = \sqrt{p_1}\ket{A}\otimes\ket{\alpha} + \sqrt{p_2}\ket{B}\otimes\ket{\beta}.
\end{equation}
Then, one sees that $\rho$ is the reduced density matrix found by
tracing out the new degrees of freedom. Let us emphasize
\emph{purification is a formal mathematical operation, and not a
  dynamical process}. In particular, for us, the new kets that we
direct producted into the Hilbert space do not correspond to any
physical degrees of freedom.\footnote{If one does treat the new
  degrees of freedom as physical in the final state, then one is
  considering a remnant scenario.} Roughly speaking, then, we can
imagine purifying each of the $\rho_i$s by enlarging the Hilbert
space, and then the evolution in this enlarged Hilbert space would be
unitary.

As it turns out, any quantum operation from say $\rho_0$ to $\rho_n$
may be written in the following way:
\begin{equation}
\rho_n = \tr_{\text{aux}}[U (\rho_{\text{aux}}\otimes \rho_0)U^\dg],
\end{equation}
for a unitary transformation $U$ acting on some auxillary degrees of
freedom as well as the physical degrees of freedom. In particular, if
$\rho_0$ acts on a $d$-dimensional Hilbert space, then we need
introduce at most a $d^2$-dimensional auxillary Hilbert space to write
the most general quantum operation in this form~\cite{nielsen}. Thus
for our $n$-qubit system, we need only introduce $2n$ auxillary qubits
to capture the most general evolution of density matrices.

\begin{figure}[ht]
\subfloat[][initial]{
\includegraphics[scale=.75, viewport=0 -24 90 68]{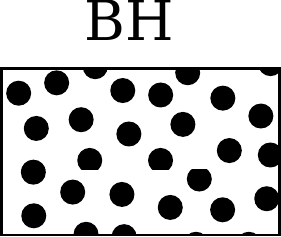}
\label{fig:initial-H}
}
\subfloat[][intermediate]{
\includegraphics[scale=.75]{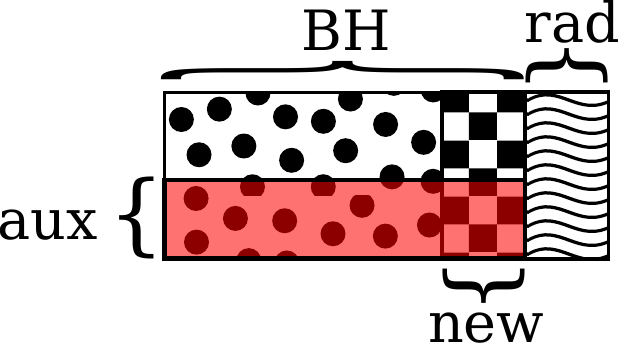}
\label{fig:int-H}
}
\subfloat[][final]{
\includegraphics[scale=.75, viewport=-10 -24 241 50]{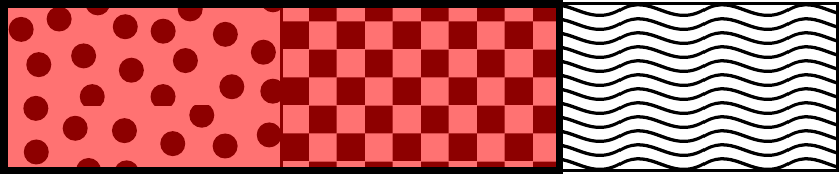}
\label{fig:final-H}
}
\caption{Here we illustrate the black hole evolution with auxillary
  degrees of freedom. At first we have only the initial black hole
  degrees of freedom represented by dots. At later stages in the
  evolution, we introduce some new degrees of freedom that
  corresponding to the infalling negative energy particles in the
  Hawking process (squares), as well as outgoing radiation (wavy
  lines). We have thus increased the size of the Hilbert space, but
  this should be thought of only as a convenient way to parametrize
  potentially nonunitary evolution. To return to a fixed dimension
  model, one needs to trace out auxillary degrees of freedom. In
  general, at intermediate steps, it is not clear which degrees of
  freedom should be thought of as auxillary, so we illustrate just one
  possibility. On the other hand, it is unambiguous that in the final
  state \emph{all} of the black hole degrees of freedom (dots and
  squares) are auxillary if there are no remnants.}
\label{fig:H}
\end{figure}

In this language, we can think about the semiclassical Hawking
evolution in the following way. We start with $n$ black hole qubits
initially in a pure state. We imagine that $n$ might be roughly given
by the entropy of the black hole.\footnote{If we want the initial
  state to model the matter just before it collapses into a black hole
  this may not be a good assumption, since in a suitably fine-grained
  description the number of degrees of freedom available to ordinary
  matter is parametrically smaller than the entropy of the black
  hole~\cite{Giddings:2009gj, 'tHooft:1993gx}. We do not concern
  ourselves with this issue here, but the author is grateful to
  S.~Giddings for pointing this out.} At the first time step, a pair
of qubits is created at the horizon in an entangled state,
\begin{equation}
\frac{1}{\sqrt{2}}(\ket{\hat{0}}\ket{0} + \ket{\hat{1}}\ket{1}).
\end{equation}
The zero represents no particle and the one represents a particle. We
refer the reader to~\cite{Mathur:2009hf, Mathur:2010kx} for a thorough
discussion on the origin of this description; see
also~\cite{Giddings:2011ks}. We have now added two new qubits to the
system, increasing the size of our Hilbert space. At each time step a
new entangled pair is produced in the above state, and the Hilbert
space keeps increasing in size.  By the end of the evaporation
process, on the $n$th step, we have added $2n$ qubits to the initial
$n$ qubits for a total $3n$ qubits.  Since the black hole has
completely evaporated and there are only the $n$ physical qubits of
radiation, the remaining $2n$ hatted qubits should be interpreted as
auxillary degrees of freedom as in the above discussion. Since
presumeably the total number of physical degrees of freedom should
remain fixed at $n$ qubits, at the $i$th step we have $2i$ auxillary
(hatted) qubits, but the semiclassical analysis does not make any
clear identification of the auxillary qubits at intermediate stages in
the evaporation. It \emph{is} clear, however, that by the end of the
evaporation process all of the black hole (hatted) qubits must be
auxillary. Specifying the auxillary subspace (in combination with
giving the internal dynamics) corresponds to taking into account back
reaction on the geometry.

The Hilbert space is illustrated in Figure~\ref{fig:H}, where the
intermediate state is shown with some highlighted auxillary degrees of
freedom. In the figure, the highlighted region contains some of the
inital matter qubits (circles) and some of the new infalling qubits
(squares); this represents one possibility. One could also consider
cases where for the first $n/2$ steps only the inital matter is
auxillary, for example. The parameterization of the auxillary degrees
of freedom at intermediate steps should be considered part of a model,
so that one may trace out the auxillary degrees of freedom to arive at
a fixed-dimensional Hilbert space description. Since we are mostly
interested in the final state, where the auxillary space is
unambiguous, we do not always specify the auxillary degrees of
freedom. In the cases where the evolution is unitary, it should be
clear what the auxillary degrees of freedom are. Although nothing
profoundly new has been said here, we hope this discussion may help
clarify potential confusions\footnote{As a specific example,
  Reference~\cite{Czech:2011wy} raises the issue of an ever-enlarging
  Hilbert space as potential issue in the analysis
  of~\cite{Mathur:2009hf}.}  regarding~\cite{Mathur:2009hf,
  Mathur:2010kx}, and other investigations~\cite{Mathur:2011wg,
  Giddings:2011ks, Czech:2011wy}.

\subsection{The General Model}\label{sec:themodel}

We are now ready to present the general class of models that we
consider.  We start with $n$ hatted qubits, and at each step we add a
hatted qubit and an unhatted qubit. The number of qubits at each step
is summarized in Table~\ref{tab:no-qubits}.

\begin{table}[ht]
\begin{center}
\begin{tabular}{c|c|c|c|c|c}
step & no. BH qubits & no. rad. qubits & total no. qubits & no. aux. qubits & state\\\hline
$0$ &  $n$   & $0$ & $n$ & 0 & $\ket{\psi_0}$\\
$1$ & $n+1$ & $1$ & $n+2$ & 2 & $\ket{\psi_1}$\\
$2$ & $n+2$ & $2$ & $n+4$ & 4 & $\ket{\psi_2}$\\
\vdots & \vdots & \vdots & \vdots & \vdots &\vdots\\
$i$ & $n+i$ & $i$ & $n+2i$ & $2i$ & $\ket{\psi_i}$\\
\vdots & \vdots & \vdots & \vdots & \vdots & \vdots\\
$n$ & $2n$ & $n$ & $3n$ & $2n$  & $\ket{\psi_n}$
\end{tabular}
\end{center}
\caption{Here we outline the discrete steps in our models. 
  At the $0$th or initial step there are $n$ black hole (BH) 
  qubits and no radiation qubits. At each step in the evolution, 
  the state is given by the ket $\ket{\psi_i}$ 
  in an enlarging Hilbert space.}\label{tab:no-qubits}
\end{table}

We model the evolution in two steps: a creation step effected by
operators $C_i$; and an internal evolution step effected by
$\hat{U}_i$ acting on the hatted qubits and $U_i$ acting on the
unhatted radiation qubits. Basis vectors at each step look like
\begin{multline}
\left\{\ket{\hat{q}_1\hat{q}_2\cdots\hat{q}_{n+i}}\ket{q_iq_{i-1}\cdots q_1}\right\}
 \xrightarrow{C_i} 
\left\{\ket{\hat{q}_1\hat{q}_2\cdots\hat{q}_{n+i}\hat{q}_{n+i+1}}\ket{q_{i+1}q_iq_{i-1}\cdots q_1}\right\}\\
\xrightarrow{\hat{U}_i\otimes U_i}
\left\{\hat{U}\ket{\hat{q}_1\hat{q}_2\cdots\hat{q}_{n+i}\hat{q}_{n+i+1}}U\ket{q_{i+1}q_iq_{i-1}\cdots q_1}\right\}.
\end{multline}
Of course, one can combine $C_i$ and $\hat{U}_i\otimes U_i$ into a
single operator, but it is useful to break up the evolution in this
way.  Also, there is some physical motivation for thinking about the
evolution in this way, since the pair creation time scale is roughly
$\sim M$, the black hole mass, while there are some conjectures that
the internal dynamics of the black hole should be as
fast~\cite{Hayden:2007cs, Sekino:2008he}. (For a 3+1 dimensional
Schwarzschild black hole, the scrambling time is speculated to be
$\sim M\log M$, but the evolution time step is $\sim M$.)

For the majority of the discussion, we focus on the $C_i$ and are
content to set $\hat{U}_i = U_i = I$. What properties should the $C_i$
satisfy?  We want $C_i$ to preserve the norm and be linear, which
means that we should require
\begin{equation}\label{eq:CdgC}
(C_i)^\dg C_i = I,
\end{equation}
where there is no sum on $i$.  Note that this is not the same as
$C_i(C_i)^\dg$ since the $C_i$ have \emph{non}square matrix
representations; the above requirement makes $C_i$ an isometric, but
nonunitary mapping. We also assume that the $C_i$ act only on the
hatted black hole qubits and not on the unhatted radiation qubits
which are far away from the pair creation site.

We can write the $C_i$ in the following form
\begin{equation}
  C_i = \ket{\varphi_1}\otimes \hat{P}_1+\ket{\varphi_2}\otimes \hat{P}_2+\ket{\varphi_3}\otimes \hat{P}_3 + \ket{\varphi_4}\otimes \hat{P}_4,
\end{equation}
where $\ket{\varphi_j}$ are an orthonormal basis for the created pair
qubits, and the $\hat{P}$'s are linear operators which act on the
hatted qubits (with implicit $i$ dependence).
Following~\cite{Mathur:2011wg, Giddings:2011ks}, we use the basis
\begin{equation}\begin{aligned}
\ket{\varphi^i_1} &= \frac{1}{\sqrt{2}}\big(\ket{\hat{0}_{n+i+1}}\ket{0_{i+1}} 
                           + \ket{\hat{1}_{n+i+1}}\ket{1_{i+1}}\big)\\
\ket{\varphi^i_2} &= \frac{1}{\sqrt{2}}\big(\ket{\hat{0}_{n+i+1}}\ket{0_{i+1}} 
                           - \ket{\hat{1}_{n+i+1}}\ket{1_{i+1}}\big)\\
\ket{\varphi^i_3} &= \ket{\hat{0}_{n+i+1}}\ket{1_{i+1}} \\
\ket{\varphi^i_4} &= \ket{\hat{1}_{n+i+1}}\ket{0_{i+1}},
\end{aligned}\end{equation}
for the newly created pair. The constraint in Equation~\eqref{eq:CdgC}
implies the following condition on the $\hat{P}$s:
\begin{equation}\label{eq:completeness}
(C_i)^\dg C_i = \hat{P}_1^\dg\hat{P}_1 + \hat{P}_2^\dg\hat{P}_2 
+ \hat{P}_3^\dg\hat{P}_3 + \hat{P}_4^\dg\hat{P}_4 = \hat{I}.
\end{equation}
Note that this defines the $\hat{P}$s as a set of generalized
measurement operators acting on the black hole Hilbert space. 

A fully specified model, then, entails
\begin{enumerate}
\item A set of $\hat{P}$s at each step $i$ that satisfy the completeness
  relation~\eqref{eq:completeness}.
\item The unitary operators $\hat{U}_i$ and $U_i$ for each $i$.
\item A clear delineation of the auxillary subspace at each step $i$.
\end{enumerate}
The last item is frequently omitted in our discussion; it should be
clear for unitary models, and it does not make significant differences
for the nonunitary models.  If one wants to acquire the
fixed-dimensional Hilbert space description, however, then one must
trace out the auxillary degrees of freedom at each step.  This gives a
very general model space that makes it easy to compare and contrast
different models of evolution.

\subsection{Physical Motivations}\label{sec:physics}

At this point, the class of models introduced above may seem fairly
abstract with little contact with the original black hole problem. Let
us review the physical motivations for this type of model as laid out
in~\cite{Mathur:2009hf} and in~\cite{Giddings:2011ks}. 

We consider an initial configuration of spherically symmetric matter
that forms a black hole, which we expect should be well described by
the Schwarzschild solution.  The model is based on the semiclassical
evolution of fields in the background of such a Schwarzschild black
hole. In order to give a Hilbert space description of evolution, it is
necessary to specify a spacelike slicing of the geometry so that we
can specify the quantum state of fields on each slice. It is important
to the arguments advanced in~\cite{Mathur:2009hf} that there exists a
``nice slicing'' of the black hole geometry~\cite{Lowe:1995ac}. This
slicing avoids the geometry's strong curvature, has sub-Planck scale
extrinsic and intrinsic curvature, and yet cuts through the initial
matter, horizon, and outgoing Hawking radiation in a smooth
way~\cite{Lowe:1995ac, Mathur:2009hf}. Thus, all quantum gravity
effects seem to be under control.

Our model should be considered an effective description of the
dynamics on the slicing. As is well known, in the presence of curved
backgrounds the quantum field theory notion of particle becomes
observer dependent. If we expand our fields on the slice into modes
inside the horizon and outside the horizon, then one finds that pairs
of particles are created inside and outside of the horizon. More
explicitly there is a Bogoliubov transformation such that the in
vacuum evolves to a state of the form $\exp(\gamma a_\text{inside}^\dg
a_\text{outside}^\dg)$ acting on the out vacuum. We reduce our problem
to an essentially two-dimensional one by expanding the modes in
spherical harmonics. From a two-dimensional perspective, each harmonic
corresponds to a different field. Then following~\cite{Hawking:1974sw,
  Giddings:1992ff}, as emphasized in this context
in~\cite{Giddings:2011ks}, we can use a set of modes that are
localized wavepackets so that we can talk about locality. This is
implicit in the discussion of~\cite{Mathur:2009hf}. Moreover, we can
truncate the Fock space to occupation numbers zero or one. We then are
effectively left with a discussion of qubits, with $\ket{0}$
representing no excitation and $\ket{1}$ representing an excitation.

In this description, a pair of particles are created roughly every $M$
in Planck units, with the outgoing particles traveling freely outward
on the slices and the ingoing particles traveling freely inward toward
the initial matter that is very far away on each slice. The pair of
particles are created entangled, as should be clear from the above
exponential. This all suggests an effective, discrete time evolution
with~\cite{Mathur:2009hf, Mathur:2010kx, Mathur:2011wg, Mathur:2011uj,
  Giddings:2011ks}
\begin{equation}
\hat{P}_1 = \hat{I} \qquad \hat{P}_{2,3,4} = 0\qquad
\hat{U}=U = I.
\end{equation}
Because the particles are well-separated on the slice, we do not
expect interparticle interactions to be significant which is
represented by the choice $\hat{U}=U=I$. We call this point in model
space, the Hawking model. Location of the particles on the slice can
then be read in the following way from the states. Consider, for
illustrative purposes, $n=3$ with a state of the form
\begin{equation}
  \ket{\hat{q}_1\hat{q}_2\hat{q}_3}_\text{initial}\ket{\hat{1}_4\hat{0}_5}_\text{infalling}\ket{0_2 1_1}_\text{outgoing}.
\end{equation}
The first three hatted qubits represent the initial infalling matter.
Note that in the Hawking model this matter plays no role in the
evolution. In general, we imagine that we find some qubit description
of the initial matter, the details of which are irrelevant to our
concerns here. The next two hatted qubits represent the infalling
Hawking radiation as it travels inward on the slices. We see that on
the first time step, a particle was emitted but not on the second time
step. The two unhatted bits represent the outgoing radiation, so the
above implies an outgoing particle was emitted on the first time step
and then no particle on the second time step. We have written the
qubits in the above order so that reading from left to right loosely
corresponds to traveling outward on the slice. This allows us to talk
about a coarse form of locality~\cite{Mathur:2009hf, Giddings:2011ks}.
Note that in the Hawking model the above state would be superposed
with several other direct product states.

As discussed in the Introduction and for this context
in~\cite{Mathur:2009hf, Mathur:2011uj, Mathur:2010kx,
  Giddings:2011ks}, the above semiclassical description of Hawking
evaporation is incomplete.  In particular, one expects quantum gravity
effects, backreaction, and interactions to play a role. Because of the
nature of the nice slicing and the low curvature at the horizon,
however, one generally expects all of these corrections to be small.
That is to say, on the pair creation time scale, one expects from
naive estimates that the dynamics be $\epsilon$ away from the above.
This expectation is in considerable tension with the expectation that
the dynamics are unitary.  For instance, we might introduce a set of
$\hat{P}$s that act on the last emitted ingoing particle. This would
suggest that the horizon is not effectively the vacuum and still
``remembers'' the previous emission.  One might also allow some mild
nonlocal interactions inside the black hole via some nearest neighbor
$\hat{U}_i$s. Or, motivated by holography and fast
scrambling~\cite{Hayden:2007cs, Sekino:2008he, Susskind:2011ap}, one
might consider general $\hat{U}_i$s, in which case one gives up all
notions of locality on the slice inside the black hole. The
distinction between the initial matter and the infalling particles is
also lost.  Allowing general internal dynamics does not affect the
argument of~\cite{Mathur:2009hf} and its generalization in
Section~\ref{sec:bound}, which only relies on the pair creation,
$\hat{P}_i$s, being close to the Hawking model. We will always use
$U=I$, since there is no physical motivation to consider strong
interactions among the outgoing radiation. (Such corrections would
also be irrelevant to Mathur's bound on the entanglement entropy.)
This should become clear after we examine some examples.

\section{Examples}\label{sec:models}

In this section, we highlight some special points in model space that
may be of interest and/or were discussed in the recent literature. One
of the main results of this paper is the one-parameter family of
models presented in Section~\ref{sec:one-par} that continuously
deforms the Hawking pair production in Section~\ref{sec:hawking} into
the unitary evolution in Section~\ref{sec:G2}; however, it is also
useful to write the various models in a common form, so that the
similarities and differences are manifest. This is especially true
when one wants to compare unitary models to nonunitary models. We
start with the canonical Hawking evaporation model. This should be
thought of as the baseline model to which all other models should be
compared.

\subsection{The Hawking Model}\label{sec:hawking}

The standard Hawking evaporation corresponds to creating a new pair in
the state $\ket{\varphi_1}$ irrespective of the state of the system,
as discussed at length in Section~\ref{sec:physics}.  Thus, it can be
written as
\begin{equation}\label{eq:hawking}
C^{H}_i = \ket{\varphi_1}\otimes \hat{I}\qquad \hat{U}=U = I,
\end{equation}
and so we can write the $\hat{P}$s as
\begin{equation}
\hat{P}_1 = \hat{I} \qquad \hat{P}_{2,3,4} = 0.
\end{equation}
In~\cite{Mathur:2009hf}, Mathur showed that if the created pair is at
most $\epsilon$ away from $\ket{\varphi_1}$, then the entanglement
entropy of the radiation continues to grow with each step and thus the
final state is mixed and unitarity is lost. In this language, the
bound shows that if the $\hat{P}$s are small deformations from the
above, then the final state will be mixed and the evolution will not
be unitary. In the sequel, we demonstrate this quite explicitly.

\subsection{A Burning Paper Model}\label{sec:paper}

Here, we present a unitary ``burning paper'' model that is equivalent
to the one given in~\cite{Mathur:2011wg}:\footnote{Throughout the
  discussion, the reader may assume that the identity acts on any
  subspaces which are not explicitly shown.}
\begin{equation}\begin{gathered}\label{eq:paper}
\hat{P}_1 = \hat{P}_2 = \left[\frac{1}{\sqrt{2}}\ket{\hat{0}\hat{0}}\bra{\hat{0}\hat{0}}
 + \frac{1}{2}\ket{\hat{1}\hat{0}}\bra{\hat{1}\hat{0}} 
 - \frac{1}{2}\ket{\hat{1}\hat{0}}\bra{\hat{0}\hat{1}}\right]_{n+i-1, n+i}\\
\hat{P}_3 = \left[\ket{\hat{1}\hat{0}}\bra{\hat{1}\hat{1}} 
  + \frac{1}{\sqrt{2}}\ket{\hat{0}\hat{0}}\bra{\hat{1}\hat{0}}
  + \frac{1}{\sqrt{2}}\ket{\hat{0}\hat{0}}\bra{\hat{0}\hat{1}}\right]_{n+i-1,n+i}\\
\hat{P}_4 = 0,
\end{gathered}\end{equation}
where the subscript on the brackets indicates that the qubits referred
to are the $(n+i-1)$th and the $(n+i)$th qubits. It should be clear that
this model is quite far from the Hawking model. This models the
creation step as
\begin{multline}
C_i = \ket{\hat{0}_{n+i+1}}\ket{1_i}\otimes 
       \left[\ket{\hat{1}\hat{0}}\bra{\hat{1}\hat{1}}
         + \frac{1}{\sqrt{2}}\ket{\hat{0}\hat{0}}\bra{\hat{1}\hat{0}}
         + \frac{1}{\sqrt{2}}\ket{\hat{0}\hat{0}}\bra{\hat{0}\hat{1}}\right]_{n+i-1,n+i}\\
+ \ket{\hat{0}_{n+i+1}}\ket{0_i}\otimes 
       \left[\ket{\hat{0}\hat{0}}\bra{\hat{0}\hat{0}}
         + \frac{1}{\sqrt{2}}\ket{\hat{1}\hat{0}}\bra{\hat{1}\hat{0}}
         - \frac{1}{\sqrt{2}}\ket{\hat{1}\hat{0}}\bra{\hat{0}\hat{1}}\right]_{n+i-1,n+i}.
\end{multline}
The important property of the evolution to note is that the
$(n+i+1)$th black hole qubit is always $\hat{0}$, as is the $(n+i)$th
qubit. Thus these two qubits are ``zeroed,'' and effectively
deactivated. We use the word zeroed in this sense, even if the qubit
under discussion is deactivated to a different value. (It could even
be something like $i\,\mathrm{mod}\,2$. In this situation, the
information is sometimes said to be ``bleached'' out of the state.) It
is clear, then, that these two qubits should be thought of as the
auxillary qubits at intermediate steps.  We'll discuss this a bit more
in Section~\ref{sec:unitarity}.

We also introduce some interesting internal dynamics. First, we need
to move the auxillary qubits out of the way, so that they don't affect
the next radiation step. So we first cyclically shift all of the
qubits 2 positions to the right, thus shoving the two $\hat{0}$s to
the two leftmost positions, $1$ and $2$. Then, we introduce some
dynamics for the physical degrees of freedom. We cyclically shift
\emph{only} the nonauxillary qubits to the right by one unit. This
defines $\hat{U}$ so that the model agrees with the burning paper
model studied in~\cite{Mathur:2011wg}. If we chop off the zeroed
qubits, we recover the model exactly.

The first model in~\cite{Giddings:2011ks} is in the same class of
models. It too zeroes two qubits, one of which is the newly created
black hole qubit. The main difference being that instead of the
radiation being determined by the two rightmost hatted qubits, it is
instead determined by the leftmost qubit. This model may be written as
\begin{equation}\begin{gathered}\label{eq:G1}
\hat{P}_1 = \hat{P}_2 
   = \frac{1}{\sqrt{2}}\ket{\hat{0}_{i+1}}\bra{\hat{0}_{i+1}}\otimes \hat{u}\\
\hat{P}_3 = \ket{\hat{0}_{i+1}}\bra{\hat{1}_{i+1}}\otimes \hat{u}',
\end{gathered}\end{equation}
where $\hat{u}$ and $\hat{u}$' are unitary operators acting on the
remaining hatted qubits. While it is not stated
in~\cite{Giddings:2011ks}, we should require that $\hat{u}$ and
$\hat{u}$' do not mix the first $i+1$ or the last $i$ auxillary hatted
qubits with the remaining physical qubits. The simplest case is to
take $\hat{u}=\hat{u}' = \hat{I}$. In this model, the entanglement
entropy of the radiation is always zero, which contrasts with our
expectations from~\cite{Page:1993df}.

\subsection{``Nonlocal'' Unitary Evolution}\label{sec:G2}

In~\cite{Giddings:2011ks}, Giddings presents three unitary models of
evolution. We focus on the second model that he presents. This second
model can be written in our notation as
\begin{equation}\begin{aligned}\label{eq:G2}
\hat{P}_1 &= \ket{\hat{0}_{2i+1}\hat{0}_{2i+2}}\bra{\hat{0}_{2i+1}\hat{0}_{2i+2}}\\
\hat{P}_2 &= \ket{\hat{0}\hat{0}}\bra{\hat{1}\hat{1}}\\
\hat{P}_3 &= \ket{\hat{0}\hat{0}}\bra{\hat{0}\hat{1}}\\
\hat{P}_4 &= \ket{\hat{0}\hat{0}}\bra{\hat{1}\hat{0}}
\end{aligned}\qquad \hat{U} = \hat{I},\end{equation}
where we have suppressed the $(2i+1)$ and $(2i+2)$ subscripts in all
but the first $\hat{P}$. One sees that as in the models presented in
Section~\ref{sec:paper}, two hatted qubits are zeroed at each step. In
this case, they are the $(2i+1)$th and $(2i+2)$th qubits. Thus as the
evolution progresses the hatted qubits are gradually put into a
fiducial form. By the $i$th step, the first $2i$ qubits are zeroed,
and should be thought of as auxillary. Note that the above evolution
rule breaks down on the penultimate step, when there is only one
nonauxillary qubit left. By then, we expect the black hole to be on
the Planck scale, and so we can just emit the last qubit freely.  This
model corresponds to $\theta=\frac{\pi}{2}$ in the model presented in
Section~\ref{sec:one-par}.

This model is nonlocal when one considers the model in the original
nice slicing of the black hole. In this context, the $(2i+1)$th and
$(2i+2)$th qubit are very far from the pair creation site at the
horizon. Note that this property is shared by the model in
Equation~\eqref{eq:G1}, and to a lesser extent the model in
Equation~\eqref{eq:paper}. The difference being how far away the
zeroed qubits are. One can either interpret these unitary models as
nonlocal interactions transmitting information far down the nice slice
to the horizon~\cite{Giddings:2011ks}, or in terms of fuzzball
microstates altering the state at the horizon, or as burning paper;
these information theoretic models are too crude to distinguish. This
point is discussed futher in the Conclusion.

\subsection{A Pure, but Not Invertible Model}\label{sec:G1-broken}

There are several of models of this kind. The simplest to consider is
\begin{equation}
\hat{P}_{1,2,4} = 0\qquad
\hat{P}_3 = \hat{I}\qquad \hat{U}=\hat{I}.
\end{equation}
In this model, regardless of the state of the system, the new pair is
created in the state $\ket{\varphi_3}$. The state $\ket{\varphi_3}$ is
not entangled, and thus we can think of this model as zeroing the
new black hole qubit \emph{and} the new radiation qubit. Instead of
putting the internal qubits into a fiducial form, we put the radiation
into a fiducial form. We hope that this convincingly demonstrates that
purity of the final state does not ensure unitarity.

Another interesting example of (almost) pure but not invertible
evolution is the model in Equation~\eqref{eq:G1}, with
\begin{equation}\label{eq:G1-broken}
\hat{u} = \hat{u}' = \hat{S}^{i+2, n+i}_1,
\end{equation}
where $\hat{S}^{i+2,n+1}_1$ is the operator that cyclically shifts the
$(i+2)$th through $(n+i)$th qubits to the left. For example, consider
the evolution:
\begin{equation}\begin{aligned}\label{eq:samp-run}
&\ket{\hat{0}\hat{q}_1\hat{q}_2\cdots\hat{q}_{n-2}\hat{0}}\\
\xrightarrow{C_0}
&\ket{\hat{0}\hat{0}\hat{q}_1\hat{q}_2\cdots\hat{q}_{n-2}\hat{0}}\ket{0}\\
\xrightarrow{C_1}
&\ket{\hat{0}\hat{0}\hat{0}\hat{q}_1\hat{q}_2\cdots\hat{q}_{n-2}\hat{0}}\ket{00}\\
\xrightarrow{C_2}
&\ket{\hat{0}\hat{0}\hat{0}\hat{0}\hat{q}_1\hat{q}_2\cdots\hat{q}_{n-2}\hat{0}}\ket{000}\\
\vdots
\end{aligned}\end{equation}
In the above example, the radiation is never entangled with the black
hole degrees of freedom, but its state is only determined by the first
and last qubit of the initial state. Thus, the evolution is not
unitary.  This is a potential problem with the model even as it is
defined in~\cite{Giddings:2011ks}. What happened? In essence, the
model keeps ``reading'' and zeroing the same qubit, which has the net
effect of zeroing the radiation qubits as well as the new black hole
qubits.  This illustrates that unitary evolution can be lost when
auxillary qubits mix with nonauxillary qubits.

This model is not quite pure for arbitrary initial states, since 
\begin{equation}\begin{aligned}
&\ket{\hat{1}\hat{q}_1\hat{q}_2\cdots\hat{q}_{n-2}\hat{1}}\\
\xrightarrow{C_0}
&\ket{\hat{0}\hat{1}\hat{q}_1\hat{q}_2\cdots\hat{q}_{n-2}\hat{0}}\ket{1}\\
\xrightarrow{C_1}
&\ket{\hat{0}\hat{0}\hat{0}\hat{q}_1\hat{q}_2\cdots\hat{q}_{n-2}\hat{0}}\ket{11}\\
\xrightarrow{C_2}
&\ket{\hat{0}\hat{0}\hat{0}\hat{0}\hat{q}_1\hat{q}_2\cdots\hat{q}_{n-2}\hat{0}}\ket{011}\\
\vdots
\end{aligned},
\end{equation}
which gives a pure final state, but if one considers a nontrivial
superposition of the above initial state and the one
in~\eqref{eq:samp-run} then the final state is mixed. Note that the
radiation only carries two qubits of information about the initial
state, and so the entropy of the final state is very small although
nonvanishing on some initial states. It is in this sense that we call
the evolution almost pure.

\subsection{An Impure Model}

A generic model that one constructs leads to impure evolution.  When
thinking about the different forms that $\hat{P}$s can take, a
particularly natural variation on the Section~\ref{sec:G2} model to
consider might be
\begin{equation}\begin{aligned}
\hat{P}_1 &= \ket{\hat{0}\hat{0}}\bra{\hat{0}\hat{0}}\\
\hat{P}_2 &= \ket{\hat{1}\hat{1}}\bra{\hat{1}\hat{1}}\\
\hat{P}_3 &= \ket{\hat{0}\hat{1}}\bra{\hat{0}\hat{1}}\\
\hat{P}_4 &= \ket{\hat{1}\hat{0}}\bra{\hat{1}\hat{0}},
\end{aligned}\end{equation}
where as in~\eqref{eq:G2} the above operators act on the $(2i+1)$th
and $(2i+2)$th qubits. It should be clear that this model leads to
mixed states. Note that it is also a large deformation from the
Hawking model. It is easy to see that this model is not invertible
either, by considering $\ket{\hat{0}\hat{0}}$ and
$\ket{\hat{1}\hat{1}}$ as initial states.

\subsection{Mathur--Plumberg Shift--Anti-Shift Models}\label{sec:shift}

In~\cite{Mathur:2011wg}, Mathur and Plumberg present several models.
We can write their ``Model A,'' in the following way. Let $\hat{T}_j$
be the operator that cyclically shifts \emph{only} the newly created
(not the first $n$) hatted qubits to the right by $j$.  Then, the
model is of the form
\begin{equation}
\hat{P}_1 = \lambda_1\hat{T}_1\qquad \hat{P}_2 = \lambda_2\hat{T}_{-1}
\qquad \hat{P}_{3,4}=0,
\end{equation}
where the completeness relation requires
\begin{equation}\label{eq:model-A-constraint}
|\lambda_1|^2 + |\lambda_2|^2 = 1.
\end{equation}
More generally, we can replace $\hat{T}_1$ and $\hat{T}_{-1}$ by other
unitary transformations that act on the hatted qubits. The smallness
of the corrections to Hawking is loosely determined by $\lambda_2$;
see~\cite{Mathur:2011wg} for more details and numerical results. 

The ``Model B'' of~\cite{Mathur:2011wg} may be written as
\begin{equation}
\hat{P}_1 = \lambda_1\,\hat{T}_1\otimes \ket{\hat{1}_{i+1}}\bra{\hat{1}_{i+1}}
+ \hat{I}\otimes\ket{\hat{0}_{i+1}}\bra{\hat{0}_{i+1}}
\quad \hat{P}_2 = \lambda_2\,\hat{T}_{-1}\otimes\ket{\hat{1}_{i+1}}\bra{\hat{1}_{i+1}}\quad  \hat{P}_{3,4}=0,
\end{equation}
where one can confirm that the completeness relation imposes the same
constraint~\eqref{eq:model-A-constraint}. 

Note that neither of the above models zeroes any qubits, and so one
does not expect information about the initial matter to be transmitted
out in the radiation. 

\subsection{Mathur ``Ising'' Model}

In~\cite{Mathur:2010kx}, Mathur presents a model that can be mapped
onto the one-dimensional Ising model and thus solved analytically. The
model can be written in the form
\begin{equation}
\hat{P}_1 = \lambda_1 \left[\ket{\hat{0}}\bra{\hat{0}} 
          +  \ket{\hat{1}}\bra{\hat{1}}\right]_{n+i}\qquad
\hat{P}_2 = \lambda_2 \left[\ket{\hat{0}}\bra{\hat{0}} 
          - \ket{\hat{1}}\bra{\hat{1}}\right]_{n+i}\qquad
 \hat{P}_{3,4}=0,
\end{equation}
where $\lambda_1$ and $\lambda_2$ are related to the parameters $a$
and $b$ in~\cite{Mathur:2010kx} via
\[
\lambda_1 = \frac{e^a + e^b}{2}\qquad \lambda_2 = \frac{e^a-e^b}{2}.
\]
The above rule is not valid for the first step, for which we use the
Hawking model ($\lambda_1 =1$ and $\lambda_2 = 0$).  The parameters
$\lambda_1$ and $\lambda_2$ are subject to the same constraint as
before
\begin{equation}
|\lambda_1|^2 + |\lambda_2|^2 = 1.
\end{equation}
In fact, this model is qualitatively the same as the model in
Section~\ref{sec:shift}: both set $\hat{P}_1$ and $\hat{P}_2$ to 
unitary transformations, and $\hat{P}_3=\hat{P}_4=0$.  The expression
for the final state entropy can be written in the
form~\cite{Mathur:2010kx}
\begin{equation}
S = n\log 2 - (n-1)\left[ae^{2a} + be^{2b}\right].
\end{equation}
Let us note that for no value of $a$ and $b$ that solves the
completeness relation does this vanish. Thus, no matter how large the
corrections are in this model, unitarity is lost; however, one
\emph{can} set $\lambda_1 = \lambda_2$ in which case the entanglement
entropy of the radiation is $\log 2$ for all time. In this case, the
pair creation (after the first step) is given by
\begin{equation}
C_i = \ket{\hat{0}_{n+i+1}}\ket{0_i}\otimes\ket{\hat{0}_{n+i}}\bra{\hat{0}_{n+i}}
     +\ket{\hat{1}_{n+i+1}}\ket{1_i}\otimes\ket{\hat{1}_{n+i}}\bra{\hat{1}_{n+i}}.
\end{equation}
One sees that after the first step, no further entanglement between
the hatted and unhatted qubits is generated. One might think there
would be more entanglement generated when the above acts on qubits
which are in a superposition of $1$ and $0$, but since the above only
acts on previously created pairs we do not have to worry about that
issue. One can imagine that the model breaks down when the black hole
becomes very small and the last qubit is emitted freely, so this model
is effectively pure in this case. Let us note that keeping the
entanglement entropy of the radiation constant is in stark contrast
with expectations we have from Page~\cite{Page:1993df}; there is no
characteristic rise and fall of the entanglement entropy of the
radiation. Furthermore, the final state consisting entirely of
radiation is \emph{completely} independent of the initial matter that
formed the black hole; the evolution is (almost) pure but far from
invertible like the model discussed in Section~\ref{sec:G1-broken}.
This model, while interesting to consider, never leads to unitary
evolution, even when one considers arbitrarily large deformations from
the Hawking point in model space.

\section{Review and Generalization of Mathur's Argument}\label{sec:bound}

In~\cite{Mathur:2009hf}, Mathur argues that small corrections to the
pair creation process are insufficient to restore unitarity. More
specifically, he demonstrates that small corrections don't accumulate.
Let us define $S_i$ as the entanglement entropy of the radiation with
the rest of the system at step $i$. If the black hole evaporation
process is unitary, then in the limit of large $n$ we expect $S_i$ to
rise linearly with $i$ until about the halfway point, $i=n/2$, and
then rapidly turn over and fall linearly to zero on the final
step~\cite{Page:1993wv}. On the other hand, if one uses the Hawking
model of evolution $C^H$, then one sees that $S_i$ increases by $\log
2$ at each time step:
\begin{equation}
S^{\text{Hawking}}_i = i \log 2;
\end{equation}
there is no turnover.  This is also the maximum entropy that is
possible for the $i$ radiation qubits, which indicates that the
radiation carries no information about the initial state.

The central insight in Mathur's argument~\cite{Mathur:2009hf} is that
the marginal increase in entanglement entropy,
\begin{equation}
\Delta S_i = S_{i+1}-S_i,
\end{equation}
varies smoothly with small deformations away from the Hawking model,
and thus a large deformation is needed to make $\Delta S$ negative if
one starts from the Hawking model's $\log 2$. In~\cite{Mathur:2009hf},
only a $\hat{P}_2$-type deformation to the Hawking model was
explicitly considered; however, we demonstrate below that the the
argument generalizes to all deformations in the model space.

In particular, we claim in the class of models discussed in this
paper, if
\begin{equation}
\|C_i - C_i^H\|<\veps< 1,
\end{equation}
then
\begin{equation}\label{eq:my-bound}
\Delta S_i \geq \log 2 - k_\veps,
\end{equation}
where $k_\veps$ is parametrically small, positive, and vanishes as
$\veps$ goes to zero. In fact, we demonstrate below that $k_\veps$
behaves as no worse than $k_\veps\sim -9\veps\log\veps$ as 
$\veps$ approaches zero.  Above, $\|\cdot\|$ is the operator norm. For
an operator $O$, $\|O\|$ is the square root of the largest eigenvalue
of $O^\dg O$.\footnote{In fact, the result is unchanged by using any
  other norm that is compatible with the Hilbert space norm.}

The proof follows that in~\cite{Mathur:2009hf} with some minor
modifications.  To begin, we use strong subadditivity of (von Neumann)
entanglement entropy.  Throughout, without loss of generality, we set
$\hat{U}=U=I$ since they do not affect the entanglement entropies. Let
us consider the $(i+1)$th state
\begin{equation}
\ket{\psi_{i+1}} = C_i\ket{\psi_i}
\end{equation}
Let $R_i$ denote the first $i$ emitted radiation qubits, $r$ denote
the $(i+1)$th emitted radiation qubit, $B_i$ denote the first $n+i$
black hole qubits, and $b$ denote the $(n+i+1)$th black hole qubit. Then,
strong subadditivity implies
\begin{equation}
S(R_i\cup r)+S(r\cup b)\geq S(R_i) + S(b).
\end{equation}
By definition, $S(R_i\cup r)$ is simply $S_{i+1}$, whereas since the
$C_i$ acts trivially on emitted radiation $S(R_i) = S_i$. Thus, we can
write the above as
\begin{equation}
\Delta S_i \geq S(b) - S(r\cup b).
\end{equation}
Note that for the Hawking model $S(b) = \log 2$ and $S(r\cup b) = 0$,
and the bound is saturated.

Now, we need to use the condition on $C_i$ to place bounds on $S(b)$
and $S(r\cup b)$. The operator norm is compatible with the Hilbert
space norm, which means
\begin{equation}
\|(C_i - C^H)\ket{\lambda}\|\leq \|C_i - C^H\|\,\|\ket{\lambda}\|
\end{equation}
for all $\ket{\lambda}$. Applying this to $\ket{\psi_i}$ gives the
condition
\begin{equation}\label{eq:ket-bound}
\|\ket{\psi_{i+1}} - C_i^H\ket{\psi_i}\| < \veps.
\end{equation}
We may write the two kets in the form
\begin{equation}
\ket{\psi_{i+1}} = \alpha_1 \ket{\varphi_1}\ket{\Lambda_1}
+ \alpha_2 \ket{\varphi_2}\ket{\Lambda_2}
+ \alpha_3 \ket{\varphi_3}\ket{\Lambda_3}
+ \alpha_4 \ket{\varphi_4}\ket{\Lambda_4},\qquad
C_i^H\ket{\psi_i} = \ket{\varphi_1}\ket{\Lambda_0},
\end{equation}
where the $\ket{\Lambda_i}$ are normalized, but not necessarily
orthogonal kets in the $n+2i$ qubit space $R_i\cup B_i$. Of course
normalization demands that
\begin{equation}
|\alpha_1|^2 + |\alpha_2|^2+|\alpha_3|^2+|\alpha_4|^2 = 1.
\end{equation}
One can show that the condition~\eqref{eq:ket-bound} implies that
\begin{equation}
\Re\big(\alpha_1 \braket{\Lambda_0|\Lambda_1}\big) > 1-\frac{\veps^2}{2},
\end{equation}
and since $|\braket{\Lambda_0|\Lambda_1}|\leq 1$, we see
\begin{equation}
1-\frac{\veps^2}{2}<|\alpha_1|\leq 1.
\end{equation}
This, in turn, implies that
\begin{equation}\label{eq:delta-def}
1-\delta^2 <|\alpha_1|^2<1,\qquad |\alpha_2|,|\alpha_3|,|\alpha_4|<\delta = \veps\sqrt{1-\frac{\veps^2}{4}},
\end{equation}
where we have defined $\delta$ for convenience. Note that $\delta$ is
what should be directly compared with $\epsilon$
in~\cite{Mathur:2009hf}.

Let us use the above to place an upper bound on $S(r\cup b)$. The
corresponding reduced density matrix is given by
\begin{equation}
\rho_{rb} = \begin{pmatrix}
|\alpha_1|^2 & \alpha_1\alpha_2^*\braket{\Lambda_2|\Lambda_1} 
  &  \alpha_1\alpha_3^*\braket{\Lambda_3|\Lambda_1} 
  &  \alpha_1\alpha_4^*\braket{\Lambda_4|\Lambda_1}\\
\alpha_1^*\alpha_2\braket{\Lambda_1|\Lambda_2} & |\alpha_2|^2 
  & \alpha_2\alpha_3^*\braket{\Lambda_3|\Lambda_2} 
  & \alpha_2\alpha_4^*\braket{\Lambda_4|\Lambda_2}\\
\alpha_1^*\alpha_3\braket{\Lambda_1|\Lambda_3} 
  & \alpha_2^*\alpha_3\braket{\Lambda_2|\Lambda_3} & |\alpha_3|^2 
  & \alpha_3\alpha_4^*\braket{\Lambda_4|\Lambda_3}\\
\alpha_1^*\alpha_4\braket{\Lambda_1|\Lambda_4} 
  & \alpha_2^*\alpha_4\braket{\Lambda_2|\Lambda_4} 
  & \alpha_3^*\alpha_4\braket{\Lambda_3|\Lambda_4} & |\alpha_4|^2 
\end{pmatrix}.
\end{equation}
We can now use Audenaert's optimal generalization~\cite{audenaert} of
Fannes' inequality, which places a bound on the difference in entropy
of two $d$-dimensional density matrices $\rho$ and
$\sigma$.\footnote{One could instead use Fannes' inequality for a
  weaker bound with stronger restrictions on $\veps$.} Let $T$ be the
trace distance between $\rho$ and $\sigma$, then~\cite{audenaert}
\begin{equation}\label{eq:fannes}
|S(\rho) - S(\sigma)|\leq T\log (d-1) - T\log T-(1-T)\log(1-T),
\end{equation}
where the trace distance is defined as
\begin{equation}
T  = \frac{1}{2}\tr\left[\sqrt{(\rho-\sigma)^\dg(\rho-\sigma)}\right],
\end{equation}
or one-half the sum of the absolute value of the eigenvalues of
$\rho-\sigma$. Note that the above definition of the trace distance
differs by a factor of $2$ from some references. With the above
normalization $0\leq T\leq 1$ for all unit-trace density matrices
$\rho$ and $\sigma$.  In this case we consider the $\sigma$ to be the
density matrix with $|\alpha_1| = 1$, and $\rho$ to be $\rho_{rb}$.
Next, we may use Gershgorin's circle theorem to bound the eigenvalues,
$\lambda$, and therefore the trace distance, $T$.  Gershgorin's
theorem tells us all the eigenvalues must lie in the union of discs in
the complex plane centered on the diagonal entries with radii given by
the sum of the absolute value of off-diagonal entries for each row.
For instance the first row gives a disc
\begin{equation}
D_1:\quad |\lambda -(|\alpha_1|^2-1)| \leq 
|\alpha_1|\,|\alpha_2|\, |\braket{\Lambda_2|\Lambda_1}|
  +  |\alpha_1|\,|\alpha_3|\,|\braket{\Lambda_3|\Lambda_1}| 
  +  |\alpha_1|\,|\alpha_4|\,|\braket{\Lambda_4|\Lambda_1}|
\end{equation}
Applying our inequalities on the components and the hermiticity of
$\rho-\sigma$ (and therefore reality of its spectrum), we can find an
interval that must include any eigenvalues in the above disc:
\begin{equation}
I_1= (-3\delta-\delta^2, 3\delta).
\end{equation}
One finds the remaining three rows can be encompassed by the interval
\begin{equation}
I_2 = (-\delta-2\delta^2, \delta+3\delta^2),
\end{equation}
and so all eigenvalues must satisfy
\begin{equation}
|\lambda| < 3\delta+\delta^2.
\end{equation}
Thus, we may conclude that the trace distance must satisfy
\begin{equation}
T < 2(3 \delta + \delta^2),
\end{equation}
and therefore 
\begin{equation}\label{eq:x1}
S(r\cup b) \leq  -x_1
               \log\left(\tfrac{1}{3}x_1\right)
            - (1-x_1)\log(1-x_1)\qquad
           x_1 =\min\big(\tfrac{3}{4},\, 2(3 \delta + \delta^2)\big).
\end{equation}
The $3/4$ comes by finding critical point of the right-hand side
of~\eqref{eq:fannes} as a function of $T$.

The bound on $S(b)$ can be derived in analogous fashion. The state
$\ket{\psi_{i+1}}$ may be written out as
\begin{equation}
\begin{gathered}
\ket{\psi_{i+1}} = \ket{\hat{0}}\ket{\chi_0}+\ket{\hat{1}}\ket{\chi_1}\\
\ket{\chi_0} = \frac{\alpha_1}{\sqrt{2}}\ket{0}\ket{\Lambda_1}
              +\frac{\alpha_2}{\sqrt{2}}\ket{0}\ket{\Lambda_2}
              +\alpha_3\ket{1}\ket{\Lambda_3}\\
\ket{\chi_1} = \frac{\alpha_1}{\sqrt{2}}\ket{1}\ket{\Lambda_1}
              -\frac{\alpha_2}{\sqrt{2}}\ket{1}\ket{\Lambda_2}
              +\alpha_4\ket{0}\ket{\Lambda_4}.
\end{gathered}
\end{equation}
The reduced density matrix can be written as
\begin{equation}
\rho_b = \begin{pmatrix}
\braket{\chi_0|\chi_0} & \braket{\chi_1|\chi_0}\\
\braket{\chi_0|\chi_1} & \braket{\chi_1|\chi_1}
\end{pmatrix}.
\end{equation}
As before we can place bounds on the above components
\begin{equation}
\begin{aligned}
\left|\braket{\chi_0|\chi_0}-\frac{1}{2}\right|&<\delta + \frac{\delta^2}{2}\\
\left|\braket{\chi_1|\chi_1}-\frac{1}{2}\right|&<\delta + \frac{\delta^2}{2}\\
|\braket{\chi_1|\chi_0}|&<\sqrt{2}(\delta + \delta^2).
\end{aligned}
\end{equation}
One can once again use the Fannes--Audenaert inequality along with
Gershgorin's theorem to bound $S(b)$, where $\sigma$ is the
$\alpha_1=1$ density matrix, $I/2$. One finds the trace distance
satisfies
\begin{equation}
T < (1+\sqrt{2})\delta + (\tfrac{1}{2}+\sqrt{2})\delta^2,
\end{equation}
and this gives
\begin{equation}\label{eq:x2}
S(b) \geq\log 2 + x_2\log x_2 + (1-x_2)\log(1-x_2)\qquad
 x_2 = \min\big(\tfrac{1}{2},\,(1+\sqrt{2})\delta + (\tfrac{1}{2}+\sqrt{2})\delta^2\big).
\end{equation}
Finally, this allows us to write
\begin{equation}\label{eq:k}
k_\veps = -x_1\log (\tfrac{1}{3}x_1) - x_2\log x_2 -(1-x_1)\log(1-x_1)-(1-x_2)\log(1-x_2),
\end{equation}
where recall $x_1$ is defined in Equation~\eqref{eq:x1}, $x_2$ in
Equation~\eqref{eq:x2}, and $\delta$ in Equation~\eqref{eq:delta-def}.
For asymptotically small $\veps$, $k_\veps\sim
-(7+\sqrt{2})\veps\log\veps$; and in fact for small but finite
$\veps$, $k_\veps< -9\veps\log\veps$. Numerically, one finds that
$k_\veps$ first surpasses $\log 2$, thus allowing the entanglement
entropy to decrease for $\veps \approx .02$. Furthermore, one finds
$k_\veps$ reaches $2\log 2$, thus allowing the maximal marginal
decrease of entanglement entropy for $\veps \approx .05$. For larger
$\veps$, the inequality with $k_\veps$ given in~\eqref{eq:k} places no
restriction on the marginal change in entanglement.

Since we are not especially interested in making the tightest possible
bound or even the above numerical values, we may as well write the
bound in slightly less unwieldy form,
\begin{equation}
\Delta S_i \geq \log 2 + 9\veps\log\veps\qquad \veps\ll 1.
\end{equation}
This establishes the claim. Let us note that the above bound's
asymptotic behavior is weaker than the inequality derived
in~\cite{Mathur:2009hf}\footnote{An equivalently strong bound here
  would be $k_\veps = 2\delta$.} as a consequence of using more
general arguments to include arbitrary perturbations.  On the other
hand, the result presented here is stronger in the sense that
\cite{Mathur:2009hf} finds only a leading order result valid to order
$O(\epsilon^2)$ whereas the bound~\eqref{eq:my-bound} with $k_\veps$
given in~\eqref{eq:k} is valid for finite $\veps\in(0,1)$.  The above
bounds could possibly be strengthened with more work;\footnote{For
  example, one might make progress by direct computation of $S(r\cup
  b)$ and $S(b)$ as was performed in~\cite{Mathur:2009hf}, but this
  would involve solving an eigenvalue problem for a four-dimensional
  density matrix with arbitrary coefficients.}  however, that is
irrelevant to the basic claim that small corrections to the low energy
pair creation process cannot restore unitarity.

\section{Requirements for Unitarity}\label{sec:unitarity}

While it is interesting to think about the different kinds of
evolution that one could have, perhaps the most interesting question
to ask is what kinds of models are unitary or, equivalently what sorts
of corrections to the Hawking evolution can restore unitarity.  Above
we see that small corrections to the evolution \emph{cannot} restore
unitarity; this gives a necessary condition that the corrections are
\emph{large}. It would be nice to also have some sufficient
conditions, since it is clear that not every large correction one
could consider leads to unitary evolution.

In order for the evolution to be pure, we need the final state
(including the auxillary qubits) to be a direct product of the form
\begin{equation}
\ket{\psi_n} = \ket{\hat{\phi}}\otimes\ket{\chi},
\end{equation}
where $\ket{\hat{\phi}}$ is a state in the $2n$-qubit auxillary space
and $\ket{\chi}$ is the state of the physical radiation qubits. Our
first observation is that $\ket{\hat{\phi}}$ should be independent of
the initial state. Suppose that this were not true:
\begin{equation}\begin{aligned}
\ket{\phi_0^{(1)}} &\mapsto \ket{\hat{\phi}^{(1)}}\otimes\ket{\chi^{(1)}}\\
\ket{\phi_0^{(2)}} &\mapsto \ket{\hat{\phi}^{(2)}}\otimes\ket{\chi^{(2)}}
\end{aligned},
\end{equation} 
which seems fine until one considers an initial state which is a
superposition of the above two states; the final state is then mixed
when one traces out the hatted qubits. (This argument assumes that the
$\ket{\chi}$s are linearly independent so that the evolution is
invertible.) This is basically a variant of the no-cloning theorem.

The next observation is that we need the final radiation state
$\ket{\chi}$ to be a unitary transformation of the initial state
$\ket{\psi_0}$. Let $F$ be the total map from initial state to the
final state, then $F$ is a linear, isometric (norm-preserving) mapping
from $n$ hatted qubits to $2n$ hatted plus $n$ unhatted qubits. From
the above, unitarity demands that
\begin{equation}\label{eq:Funit}
  F_\text{unitary}: \ket{\hat{\psi}}\mapsto \ket{\hat{\phi}}\otimes \ket{\chi}\qquad
 \ket{\chi} = U\ket{\hat{\psi}},
\end{equation}
where $U$ in the above is a unitary transformation from the initial
$n$ hatted qubits to the final $n$ unhatted qubits, and
$\ket{\hat{\phi}}$ is fixed. The total map $F$ is just the product of
all the $C_i$s and $\hat{U}_i\otimes U_i$s. All of the hatted qubits
have to be zeroed or bleached, and the information stored in the
initial matter transferred to the radiation.

Let us think about how we can zero or bleach the hatted qubits.  We
have $n$ steps to project $2n$ qubits to a unique state with the
$\hat{P}^i$s; the unitary $\hat{U}$s clearly cannot zero qubits. At
each step, we can zero \emph{at most two qubits}. If $C$ bleaches some
subspace to a state $\ket{\hat{\alpha}}$, then it may be written as
\begin{equation}
C = \ket{\hat{\alpha}}\otimes O,
\end{equation}
for an unspecified operator $O$. If $\ket{\hat{\alpha}}$ is a
$p$-qubit subspace, then $O$ maps $n+i$ qubits to $n+i+2-p$ qubits and
must satisfy
\begin{equation}
O^\dg O = I;
\end{equation}
this is only possible if $n+i+2-p\geq n+i$, immediately implying
$p\leq 2$.

Our key observation is that the desire to zero the hatted qubits is in
tension with the completeness relation~\eqref{eq:completeness}. Since
we only have four $\hat{P}$s, at any given step the \emph{best} we can
do is project out a four-dimensional subspace, or two qubits.  It is
this tension that connects the need to zero qubits with the
requirement to have large corrections to the Hawking model. This might
help elucidate the results in~\cite{Mathur:2009hf}.  Note that this is
very much in agreement with the picture presented in
Figure~\ref{fig:H}, wherein at each stage there are two new auxillary
qubits. For the state to be pure, these auxillary qubits must be
zeroed.

Let us note that there are two different ways to zero two qubits at
each step, although the distinction is not actually that sharp when
one considers the full $C_i$s. In the burning paper model of
Section~\ref{sec:paper} we use only three $\hat{P}$s, which zero one
qubit. The three $\hat{P}$s were chosen to ensure that the newly
created $\hat{q}$ is also zeroed. The second way is illustrated in
Section~\ref{sec:G2}, in which all four $\hat{P}$s are used to zero
two old $\hat{q}$s. One can consider various unitary transformations,
however, these are the only two qualitative kinds of models that lead
to a pure radiation final state. Remember that which $\hat{P}$s get
used tell us which pair state is created at the horizon. It is
impossible to preferentially use only $\hat{P}_1$ and simultaneously
have unitary evolution.

As we saw in Section~\ref{sec:G1-broken}, it is possible for the
evolution to be pure, but not invertible. In the model in
Equation~\eqref{eq:G1-broken}, qubits that were zeroed in previous
steps mixed with nonzeroed qubits. When we zero the qubits, we are
then thinking of them as auxillary degrees of freedom that should be
erased in the operator-sum description~\eqref{eq:op-sum-rep}.  Thus,
it does not make physical sense to allow mixing with the auxillary
degrees of freedom if one wants unitary evolution.

The requirements outlined above for purity and invertibilty should
ensure unitary evolution.

\section{A One-Parameter Interpolating Model}\label{sec:one-par}

We can interpolate between the Hawking model in
Section~\ref{sec:hawking} and the unitary model in
Section~\ref{sec:G2} via
\begin{equation}\begin{aligned}\label{eq:theta-model}
\hat{P}_1 &= \cos\theta\,\hat{I} 
   + (1-\cos\theta)\ket{\hat{0}_{2i+1}\hat{0}_{2i+2}}\bra{\hat{0}_{2i+1}\hat{0}_{2i+2}}\\
\hat{P}_2 &= \sin\theta\ket{\hat{0}\hat{0}}\bra{\hat{1}\hat{1}}\\
\hat{P}_3 &= \sin\theta\ket{\hat{0}\hat{0}}\bra{\hat{0}\hat{1}}\\
\hat{P}_4 &= \sin\theta\ket{\hat{0}\hat{0}}\bra{\hat{1}\hat{0}}
\end{aligned}\qquad \hat{U}=\hat{I},\end{equation}
with $\theta=0$ giving Hawking's evolution and $\theta=\frac{\pi}{2}$
giving the unitary evolution in Equation~\eqref{eq:G2}. (Once again we
suppressed subscripts on the qubits after the first line.) We may
write
\begin{equation}
  (C - C^H)^\dg(C - C^H) = 2 \hat{I} - \hat{P}_1-\hat{P}_1^\dg
  = 2(1-\cos\theta)\big(\hat{I}-\ket{\hat{0}\hat{0}}\bra{\hat{0}\hat{0}}\big),
\end{equation}
so that one finds
\begin{equation}
\|C - C^H\| = 2|\sin\tfrac{\theta}{2}|.
\end{equation}
We clearly see that this is in accord with Mathur's argument and its
generalization in Section~\ref{sec:bound}.

\begin{figure}[htb]
\subfloat[][$\ket{\hat{0}\hat{0}\hat{0}\hat{0}\hat{0}}$]{
\includegraphics[width=5cm]{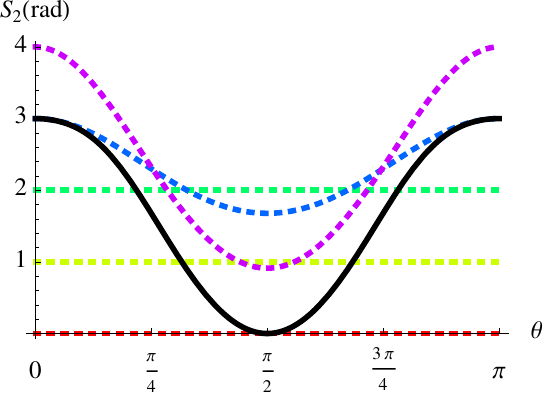}
}
\subfloat[][$\ket{\hat{0}\hat{0}\hat{0}\hat{1}\hat{1}}$]{
\includegraphics[width=5cm]{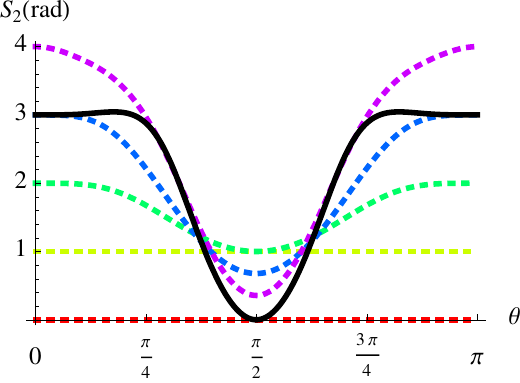}
}
\subfloat[][$\ket{\hat{1}\hat{1}\hat{1}\hat{1}\hat{1}}$]{
\includegraphics[width=5cm]{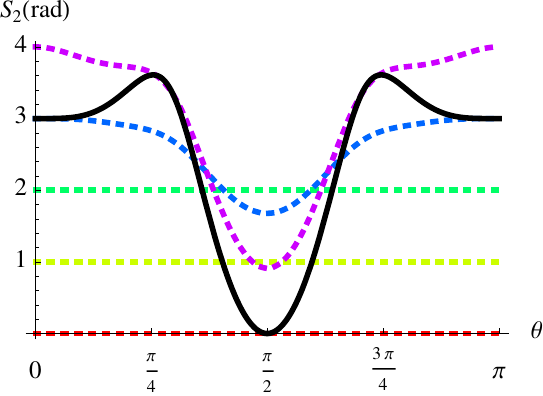}
}
\caption{Here we present the second R\'{e}nyi entropies of the
  unhatted radiation qubits as a function of $\theta$ for the model in
  Equation~\eqref{eq:theta-model} with three different initial states.
  Intermediate steps are dashed curves and the final step is solid;
  the steps are shown in the color order (red, yellow, green, blue,
  purple, black). The point $\theta=0$ corresponds to the canonical
  Hawking evolution, while $\theta=\frac{\pi}{2}$ corresponds to the
  unitary model in Equation~\eqref{eq:G2}.}\label{fig:theta}
\end{figure}

One of the main results of this paper is the above model, which
continuously connects the Hawking model to a unitary model, clearly
illustrating that they are far apart in model space. Previous efforts
to illuminate Mathur's bound~\cite{Mathur:2011wg, Mathur:2010kx},
considered different types of small corrections and showed that they
did not significantly affect the entropy of the final state; however,
they did not consider corrections that when made \emph{large} would
give a unitary ``burning paper'' type model. The above model fills
this gap. In Figure~\ref{fig:theta}, we plot the second R\'{e}nyi
entanglement entropies of the radiation as a function of $\theta$ for
three different initial states. Recall that the second R\'{e}nyi
entropy is defined as
\begin{equation}
S_2(\rho_\text{red}) = - \log\tr(\rho^2_{\text{red}}),
\end{equation}
and is a positive-definite measure of entanglement that vanishes if
and only if $\rho_{\text{red}}$ is pure. For computational purposes,
however, it is a bit more convenient than the traditional von Neumann
entropy. It should also be noted that $S_2$ gives a lower bound for
the von Neumann entropy. Moreover, when one examines the von Neumann
entropy it behaves qualitatively similarly. 

In Figure~\ref{fig:theta}, one sees that at $\theta=0$ (Hawking model)
the entropy rises by one at each step, except for the final step where
it drops down by one. This is an artifact of the way we end the
evolution, since the model breaks down on the penultimate step. We
chose to just emit the last qubit freely, which makes sense on
physical grounds since the black hole should be quite small by that
point; one can easily imagine large corrections in the final stage(s)
of black hole evaporation. At $\theta=\frac{\pi}{2}$, we have the
unitary model, and the entanglement entropy of the radiation has the
expected rise and fall. For no $\theta$ close to $\theta=0$, however,
does the entanglement entropy of the radiation fail to increase
(excluding the final step). Of course, a $5$ qubit initial state is
not realistic for the macroscopically large black holes we are
thinking about; however, it suffices to demonstrate the qualitative
behaviour which should not change as $n$ is increased.  The author was
limited by the computational power required, which grows quite rapidly
with $n$.

\section{Conclusion}\label{sec:conc}

We have presented a very general framework that provides a natural,
unifying language to compare information-theoretic models of black
hole evaporation. The framework involves describing the dynamics of
pure state vectors $\ket{\psi}$ in an ever enlarging Hilbert space.
The only constraint on the models is the completeness
relation~\eqref{eq:completeness}. In Section~\ref{sec:quops}, we
explain how to interpret this model in terms of potentially mixed
evolution in a fixed dimensional Hilbert space: one must trace out
some auxillary degrees of freedom to arrive at a dynamical equation
like in Equation~\eqref{eq:op-sum-rep}. Part of a full model, then,
involves specifying the auxillary degrees of freedom. While at
intermediate steps this can be ambiguous, by the end of the evolution
we are left only with radiation and thus any nonradiation degrees of
freedom are by default auxillary. This excludes remnants or other
scenarios where one can identify physical degrees of freedom that the
Hawking radiation is entangled with at the end of the evaporation.

In Section~\ref{sec:models}, we show how to write a number of
interesting models in our unifying notation. Many of them had been
introduced and studied previously in the literature. These models
illustrate some of the key obstructions and requirements to have
unitary evolution. 

In Section~\ref{sec:unitarity}, we discuss the requirements for
unitarity, and summarize a set of sufficient conditions. A key
backdrop to our discussion, and indeed the whole paper, is a recent
theorem~\cite{Mathur:2009hf} showing that small corrections to the
Hawking model \emph{cannot} give unitary evolution. A corollary is
that \emph{large} corrections are necessary to have unitary evolution.
As we show, this is, however, not a sufficient condition;
unsurprisingly, there are many large corrections that fail to give
unitary evolution. One interesting observation is how the unitary
requirement that internal qubits be zeroed becomes connected to
corrections to the pair creation via the completeness
relation~\eqref{eq:completeness}.  This may help elucidate the results
in~\cite{Mathur:2009hf}.

Finally, in Section~\ref{sec:one-par}, we give a one-parameter family
of models that continuously interpolates between the Hawking model and
a unitary model. In terms of this parameter, one can clearly see that
the unitary model is far from the Hawking model, thus illustrating
the theorem in~\cite{Mathur:2009hf}.

One of the key points emphasized in~\cite{Mathur:2009hf,
  Mathur:2011uj} is that the nice slicing of the Schwarzschild
solution implies at best small corrections to the Hawking model in
Equation~\eqref{eq:hawking}, and therefore a loss of unitarity
evolution. To restore unitarity, large corrections are required of the
form discussed here; however, one must show why these corrections
arise in the black hole \emph{and not} in all of our earth-based
experiments and observations. It seems quite difficult to do this,
since in the nice slice construction no geometric quantity is large.
The only quantity that seems to be large is the number of degrees of
freedom or number of particles required to form the black hole, but
this is not a basic geometric quantity.

Let us further note that for our discussion there is a factorization
of the internal dynamics and the pair creation dynamics. The internal
black hole dynamics can be as nonlocal, or scramble as rapidly as one
wants, but whether the evaporation process is unitary or not is
(modulo a few caveats mentioned in Sections~\ref{sec:models}
and~\ref{sec:unitarity}) entirely determined by the pair creation
process. Thus, the discussion in~\cite{Hayden:2007cs, Sekino:2008he,
  Susskind:2011ap, Lashkari:2011yi} is not directly relevant to our
concerns here, although it is important for better understanding black
holes. The pair creation process is localized near the horizon, where
for a large black hole the geometry suggests one can trust the
semiclassical approximation even if one might doubt its validity deep
within the black hole. On the pair creation time scale, however, it is
precisely this physics that needs an order unity
correction~\cite{Mathur:2009hf}.

One other issue that one may wish to raise in our discussion is the
issue of conservation laws~\cite{Czech:2011wy, Braunstein:2011gz}.
While we hope we have sufficiently addressed the issue of an expanding
Hilbert space as raised in~\cite{Czech:2011wy}, one may still be
concerned that we haven't discussed conservation of energy (or angular
momentum, electric charge, etc.). These issues are rebutted
in~\cite{Mathur:2011uj}. Let us note, however, by not discussing the
original spacetime physics and the resulting at most small
corrections, the fundamental issue has been totally elided. While the
results presented in~\cite{Czech:2011wy, Braunstein:2011gz} are
interesting unto themselves, they do not provide a plausible physical
mechanism to modify the pair creation process to get the dynamics they
suggest.

If, as suggested by string theory, or from other considerations, we
think black hole evaporation is a unitary process; then, the pair
creation process must not strictly adhere to the causal structure on
the Schwarzschild nice slicing. There are two obvious frameworks
(ignoring the possibility of remnants) to discuss this deviation:
fuzzballs or nonlocality.

The fuzzball proposal (see~\cite{Skenderis:2008qn, Bena:2007kg,
  Balasubramanian:2008da, Mathur:2005ai, Mathur:2005zp} for reviews)
suggests that the black hole metric is only an effective geometry that
approximates $e^{S_\text{BH}}$ microstates. The microstates differ
from each other on the horizon scale, thus large corrections to the
Hawking evolution are anticipated and information is transmitted from
local excitations.  How the fuzzball proposal relates to these qubit
models is discussed in~\cite{Mathur:2010kx, Mathur:2011uj,
  Mathur:2009hf, Mathur:2011wg}.  The main point being that since the
geometry at the would-be horizon depends on the internal state, their
is a physical mechanism to get large corrections to the pair creation
process. Since the fuzzball's interior geometry (and the whole causal
structure) is quite different from the original black hole solution in
which the nice slices were constructed, one should probably not
interpret them with the original notion of locality for the internal
qubits discussed in Section~\ref{sec:physics}. Moreover, let us note
that by adding nontrivial dynamics of the internal fuzzball structure
via a $\hat{U}$ to the model in~\eqref{eq:G2}, one can effectively
change which qubits get emitted. This is the point referred to at the
end of Section~\ref{sec:G2}.

There is one explicit family of (non-extremal) fuzzball
microstates~\cite{Jejjala:2005yu} for which one can understand the
bulk Hawking emission process. As first suggested
in~\cite{Chowdhury:2007jx}, the geometry's ergoregion
instability~\cite{Cardoso:2007ws} can be interpreted as a Bose
enhanced version of the Hawking instability for the corresponding
black hole. This explanation was justified by comparing gravitational
emission to the dual CFT emission process for both ergoregion emission
from the fuzzballs and Hawking radiation from the corresponding black
hole~\cite{Chowdhury:2007jx, Chowdhury:2008bd, Chowdhury:2008uj,
  Avery:2009tu, Avery:2009xr}.  In~\cite{Chowdhury:2007jx}, a toy
model was presented for the ergoregion emission, based on the CFT
description. The toy model consists of a set of two-level atoms that
can spontaneously emit or absorb photons. In the geometric
description, the de-exciting atoms correspond to accumulating
particles in the ergoregion that decrease the geometry's mass and
angular momentum.  Loosely, in our language, the toy model is in the
class discussed in Section~\ref{sec:paper}. To properly capture the
Bose enhancement, however, is a bit trickier.

In~\cite{Giddings:2011ks}, several unitary models of evolution (some
of which were discussed here) were presented, motivated by proposed
nonlocal physics on the Schwarzschild background. As mentioned
in~\cite{Giddings:2009ae}, it is unclear what sets the scale of the
proposed nonlocality, so as to ensure it operates in the black hole
background but not in everyday low-energy experiments. While the sorts
of models discussed here remain too crude to distinguish between
nonlocal physics or fuzzball microstates, their utility lies in their
generality, which serves to sharpen our information theoretic
understanding of black hole evaporation. One obvious task that remains
is to translate Mathur's bound and its generalization in
Section~\ref{sec:bound} into a sharper, quantitative statement
about the breakdown of the semiclassical limit of quantum gravity.
Since the entire discussion has been in a Hamiltonian framework, it
would be especially nice to have analogous bounds on the path
integral.

\begin{acknowledgments} 
  The author is grateful for comments and correspondence related to
  this work from C.~Asplund, S.~Ghosh, S.~Giddings, and S.~Mathur. The
  material here especially benefitted from discussions with S.~Kalyana
  Rama.
\end{acknowledgments}


\bibliography{qmbh}

\end{document}